\documentclass[11pt,a4paper]{article}
\usepackage{paperstyle}
\usepackage{graphicx}
\usepackage{cite}

\title{Depositing More Should Not Earn Less}
\researchsubtitle{Supplier income monotonicity and rate curve design in pooled lending markets}
\researchshorttitle{Depositing More Should Not Earn Less}
\author{%
\begin{tabular}{c@{\hspace{3em}}c}
Ahzam Khan & Cavey Cool \\
\small\href{mailto:ahzam.main@gmail.com}{ahzam.main@gmail.com} &
\small\href{mailto:caveycool@gmail.com}{caveycool@gmail.com}
\end{tabular}}
\date{October 2026}
\hypersetup{pdftitle={Depositing More Should Not Earn Less},pdfauthor={Ahzam Khan and Cavey Cool},pdfsubject={Supplier income monotonicity and rate curve design in pooled lending markets}}

\begin{document}
\maketitle

\begin{abstract}
In utilization dependent interest rate models, adding liquidity can reduce a supplier's total interest income even as the supplier's balance increases. A supplier holding a fraction $\alpha$ of final pool supply keeps income nondecreasing with a marginal deposit exactly when $\alpha(1+\eta)\leq1$, where $\eta$ is the elasticity of the borrow rate with respect to utilization. When the condition fails, the supplier can instead raise its income by withdrawing, which pushes utilization further up. On two public Drift markets, a lender held more of the pool than the curve protected in at least one hour in nine on SOL and one in thirteen on USDC. A protocol turns this into a design constraint by choosing, at each utilization, the largest supplier share it intends to protect. That choice caps the elasticity, and integrating the cap describes every admissible curve. Protecting the same share at every utilization gives a power law boundary, and protecting every fresh depositor gives a hyperbolic one. On a multiple kink USDT market, compliant repairs can lower or raise supplier income, because the constraint governs how income responds to a deposit and not the price level.
\end{abstract}

\section{Introduction}

Pooled lending protocols commonly determine interest rates as a function of utilization~\cite{compound,aave}. A standard design increases the borrow rate gradually below an optimal utilization level and much more aggressively above it. The steep branch improves incentives to restore liquidity under stress, but it also creates a supplier price impact problem, a liquidity version of market power: an additional deposit lowers utilization, which lowers the borrow rate, and the lower rate applies to the supplier's entire balance.

Consequently, a supplier may be better off withholding capital, or withdrawing it. On a lending market at high utilization, which is when the pool most needs supply, one of us found that depositing only part of the available wallet balance earned more than depositing all of it.

Existing work compares the rate rules of deployed protocols~\cite{gudgeon}, models lending pools formally~\cite{bartoletti}, and designs rates for expected lender wealth~\cite{baude-optimal} or stable utilization~\cite{agilerate}. Baude et al.\ model how a borrower's own position moves its cost~\cite{baude-leverage} and how a lender's own supply moves its rates across several markets~\cite{baude-allocation}. We ask when one market's curve makes depositing more lower a supplier's own income, and which curves rule this out.

This paper gives an exact condition under which a deposit cannot lower the supplier's own income, describes the rate curves that satisfy it, and separates what such a guarantee promises from what it costs. Section~\ref{sec:failure} works the failure through in numbers. Sections~\ref{sec:model} and \ref{sec:condition} derive the income condition, Section~\ref{sec:evidence} shows that real suppliers on two Drift markets were exposed to it, and Section~\ref{sec:family} describes the admissible family of rate curves it induces. Sections~\ref{sec:fixed} and \ref{sec:fresh} develop two protection guarantees, and Section~\ref{sec:minimal} gives a power law repair. Section~\ref{sec:multi} takes a market with several kinks and separates protection from pricing, and Section~\ref{sec:onchain} states the check a protocol would implement. The appendices hold the detailed constructions, further examples and the smooth extension.

Most of the results are formalized in Lean~4. The Lean library, the analysis code and its data are available at \url{https://github.com/0xahzam/supplier-monotonicity}, with a record of which statements are covered.

\section{The Failure Mode}
\label{sec:failure}

Consider a market with optimal utilization $O=80\%$, a borrow rate of $R_o=5\%$ at that point, and a maximum borrow rate of $R_{\max}=100\%$ at full utilization, connected by straight lines. This is a standard kinked curve~\cite{aave}, and the change in slope at the kink is large but not unusual.

\begin{figure}[htbp]
\centering
\includegraphics[width=\textwidth]{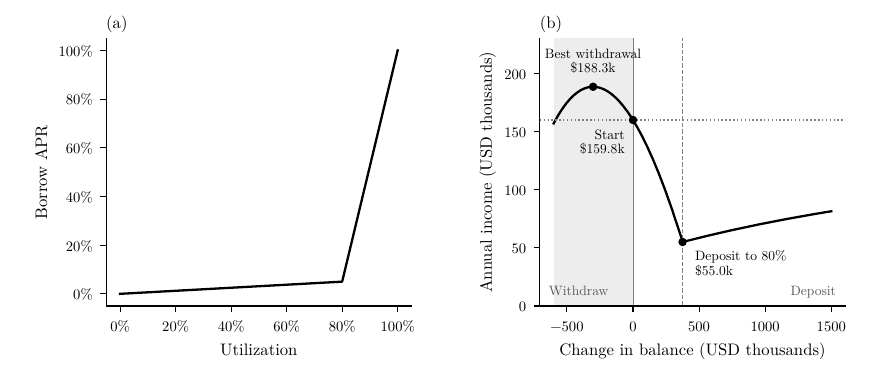}
\caption{(a) The illustrative kinked borrow rate. (b) Annualized income of a supplier initially holding $10\%$ of a \$10 million pool, with \$8.3 million of fixed borrowing, against the change in its balance. Withdrawals are shaded. The dotted line is the starting income, and the dashed line marks the deposit that returns utilization to $80\%$. The labels give income at the start, after that deposit, and at the income maximizing withdrawal.}
\label{fig:failure-mode}
\end{figure}

Let the market hold \$10 million of supply against \$8.3 million of borrowings, so utilization is $83\%$. The market is above its own optimal level, so the protocol wants more supply. A supplier that already holds \$1 million, $10\%$ of the pool, adds \$375{,}000, exactly enough to return the market to optimal utilization. Table~\ref{tab:failure} shows the result. The balance grows by $37.5\%$ and the income falls by $66\%$, because the lower rate now applies to the whole balance. Depositing past the kink recovers only part of the loss, and as Figure~\ref{fig:failure-mode} shows, the income never returns to where it started. The same supplier does better by withdrawing about \$302{,}000, which moves utilization further from the level the protocol is trying to restore.

\begin{table}[htbp]
\centering
\caption{The $10\%$ supplier at three balances. Borrowings stay at \$8.3 million. Income is annualized at the resulting state. Values in the last row are rounded.}
\label{tab:failure}
\small
\begin{tabular}{@{}lrrrrrr@{}}
\toprule
 & Balance & Total supply & $U$ & Borrow rate & Supply rate & Income \\
\midrule
Start & \$1{,}000{,}000 & \$10{,}000{,}000 & $83\%$ & $19.25\%$ & $15.98\%$ & \$159{,}775 \\
Deposit \$375{,}000 & \$1{,}375{,}000 & \$10{,}375{,}000 & $80\%$ & $5\%$ & $4\%$ & \$55{,}000 \\
Withdraw \$302{,}000 & \$698{,}000 & \$9{,}698{,}000 & $85.6\%$ & $31.5\%$ & $27.0\%$ & \$188{,}300 \\
\bottomrule
\end{tabular}
\end{table}

Two features of this example matter later. The loss comes from the sensitivity of the rate, not from the size of the deposit. The worst point is the kink, not full utilization. Both follow from the condition derived below, and neither is visible from the slope of the curve alone.

\section{Model}
\label{sec:model}

Let
\begin{itemize}
  \item $T$ denote total supplied tokens after the supplier's transaction,
  \item $B$ denote total borrowings,
  \item $b$ denote the supplier's resulting deposited balance,
  \item $U=B/T$ denote utilization,
  \item $R_b(U)$ denote the borrow rate at utilization $U$, and
  \item $R_s(U)$ denote the supply rate.
\end{itemize}

We assume the rate is nondecreasing, continuous and piecewise continuously differentiable, with $R_b(U)>0$ for $U>0$. The formulas below apply inside each smooth segment. At a kink, where the slope changes, a deposit lowers utilization, so the slope that matters is the one of the segment below. A zero base rate at $U=0$ is permitted unless a later guarantee rules it out.

Ignoring fees, aggregate borrowers pay $BR_b(U)$ in annualized interest. Distributed across $T$ supplied tokens, the supply rate is therefore
\begin{equation}
R_s(U)=\frac{BR_b(U)}{T}=UR_b(U)
\end{equation}
and the supplier's annualized interest income is
\begin{equation}
I=bR_s(U)=bUR_b(U)
\end{equation}
Income here is the annual rate at the current state, not interest already earned. Protocols usually keep a fixed share of borrower interest, often called the reserve factor. A constant share $f<1$ scales supplier income and its derivative by $1-f$ at every utilization, so it does not change the sign condition, and the analysis can ignore it. A fee that varies with utilization would have to be included in the derivative.

Write the supplier's share of final pool supply as
\begin{equation}
\alpha=\frac{b}{T}
\end{equation}
The share is measured after the deposit, at the supplier's resulting state. Equivalently $I=\alpha BR_b(U)$, so the supplier receives the share $\alpha$ of all borrower interest.

Consider an infinitesimal increase $d\delta$ in the same supplier's deposit, holding $B$ fixed during the transaction. Then
\begin{equation}
\frac{dU}{d\delta}
=\frac{d}{d\delta}\left(\frac{B}{T}\right)
=-\frac{B}{T^2}
=-\frac{U}{T}
\end{equation}
Borrowings respond to rates, but not within the transaction. A deposit settles against the current borrowings, and any borrowing that the lower rate attracts arrives later and offsets part of the loss. The condition describes that immediate effect, which is the one a supplier sees when deciding whether to deposit. How far borrowers respond afterwards is outside the claim, as in Section~\ref{sec:earnings}.

\section{Supplier Income Monotonicity}
\label{sec:condition}

Differentiating $I=bUR_b(U)$ with respect to the additional deposit gives
\begin{equation}
\begin{aligned}
\frac{dI}{d\delta}
&=\frac{d}{d\delta}\left[bUR_b(U)\right] \\
&=UR_b(U)+b\left[R_b(U)+UR_b'(U)\right]\frac{dU}{d\delta} \\
&=UR_b(U)-\frac{bU}{T}\left[R_b(U)+UR_b'(U)\right] \\
&=UR_b(U)\left[1-\frac{b}{T}\left(1+\frac{UR_b'(U)}{R_b(U)}\right)\right]
\end{aligned}
\end{equation}

\begin{proposition}[Marginal supplier income]
For a marginal deposit made while borrowings are fixed,
\begin{equation}
\boxed{
\frac{dI}{d\delta}
=UR_b(U)\left[1-\alpha\left(1+\frac{UR_b'(U)}{R_b(U)}\right)\right]
}
\end{equation}
\end{proposition}

Two competing effects sit inside that bracket: the marginal dollar earns the current supply rate $UR_b(U)$, while the fall in utilization it causes lowers the rate applied to everything the supplier already holds. The size of the second effect is measured by $UR_b'(U)/R_b(U)$, the elasticity of the borrow rate with respect to utilization,
\begin{equation}
\boxed{
\eta(U)=\frac{U}{R_b(U)}\frac{dR_b}{dU}
=\frac{UR_b'(U)}{R_b(U)}
=\frac{d\ln R_b}{d\ln U}
}
\label{eq:elasticity}
\end{equation}
The logarithmic form on the right is what Section~\ref{sec:family} works with. Substituting it into the income derivative gives the identity this paper is built on,
\begin{equation}
\boxed{
\frac{dI}{d\delta}
=UR_b(U)\left[1-\alpha(1+\eta(U))\right]
}
\label{eq:marginal}
\end{equation}
Since the prefactor is positive,
\begin{equation}
\frac{dI}{d\delta}\geq0
\quad\Longleftrightarrow\quad
\alpha(1+\eta)\leq1
\end{equation}
and the largest supplier share whose marginal deposit remains income nondecreasing is
\begin{equation}
\boxed{
\alpha_{\mathrm{safe}}(U)=\frac{1}{1+\eta(U)}
}
\end{equation}
An elasticity of $9$ protects a $10\%$ supplier, $19$ protects $5\%$, and $99$ protects $1\%$. Whether a curve is too steep for a given supplier therefore depends on its elasticity, the percentage change in rate per percentage change in utilization, and not on its slope.

A finite deposit is a chain of marginal ones. If the condition holds at every point along the way, income at the end is at least income at the start. This also holds across kinks, because income is continuous there.

The condition also explains both features of Section~\ref{sec:failure}. Above the kink the slope is $m=(R_{\max}-R_o)/(1-O)=4.75$, and on a straight segment $R_b'=m$. At the starting utilization of $83\%$ the borrow rate is $19.25\%$, so $\eta=0.83\times4.75/0.1925\approx20.5$ and $\alpha(1+\eta)\approx0.1\times21.5\approx2.15$. That is more than twice the limit, so the deposit loses money, as Section~\ref{sec:failure} found. On the steep branch elasticity is $Um/(R_o+m(U-O))$. It is largest just above the kink, where the rate is lowest. There $\eta=Om/R_o=76$, so only suppliers below $1/77\approx1.3\%$ of the pool are protected. At full utilization it falls to $m/R_{\max}=4.75$, because the same slope moves a high rate by a smaller percentage. The steep branch is therefore most dangerous at its foot rather than its top. Below the kink this curve has $\eta=1$, so the condition holds again for any share up to one half and income starts to rise. That is why the income trough in Figure~\ref{fig:failure-mode} sits exactly at the kink.

Read in the other direction, the same condition prices a withdrawal. Wherever $\alpha(1+\eta)>1$ the derivative in Equation~\ref{eq:marginal} is negative, so removing liquidity raises the supplier's income. Withdrawing lowers the supplier's share and raises utilization, which on the steep branch lowers elasticity, so $\alpha(1+\eta)$ falls until income peaks where it reaches $1$. In Section~\ref{sec:failure} this is the withdrawal of about \$302{,}000, at which $\alpha(1+\eta)=1$ exactly. So the supplier the pool most needs earns more by leaving, and leaving pushes utilization further above $O$. A withdrawal can draw only on liquidity that is not lent out, so it can never push utilization above $1$. In the Section~\ref{sec:failure} example, about \$1.7 million was unlent.

In the limit of a supplier that owns the whole pool, $\alpha=1$ and the condition requires $\eta\leq0$. Equation~\ref{eq:marginal} then becomes
\begin{equation}
\frac{dI}{d\delta}=-U^2R_b'(U)
\end{equation}
so no strictly increasing rate curve protects a sole supplier, and only a flat rate keeps its income unchanged.

The condition is the marginal revenue condition for a seller with market power~\cite{cournot,lerner}. Liquidity is the good and the supply rate is its price. Since $R_s=BR_b(B/T)/T$, adding one percent to total supply lowers the supply rate by about $1+\eta$ percent, and Equation~\ref{eq:marginal} reads $dI/d\delta=R_s[1-\alpha(1+\eta)]$, the marginal revenue of a seller holding the share $\alpha$ of that market. Withdrawing when the bracket is negative is the matching withholding of output, and the sole supplier above is the monopolist. What differs here is that the protocol sets the demand curve itself, so it can choose the elasticity that removes this incentive for every supplier below a chosen share. Sections~\ref{sec:family} to \ref{sec:multi} develop this.

\section{Evidence from Drift Markets}
\label{sec:evidence}

To see whether real suppliers end up in the position of Section~\ref{sec:failure}, we check, hour by hour, whether any supplier on a live market held a larger share of the pool than the curve protected, that is, whether its share exceeded $\alpha_{\mathrm{safe}}$. We call such a supplier \emph{exposed}, since by Equation~\ref{eq:marginal} depositing more would have lowered its own income.

We use the USDC and SOL lending markets of Drift~\cite{drift}, a trading and lending protocol on Solana, from 11 November 2024 to 1 April 2026. These are public markets, unrelated to the multiple kink example later in the paper. Drift writes an onchain record each time a market changes, including its total deposits, total borrows and the rate curve in force, which gives utilization and $\alpha_{\mathrm{safe}}$ for every hour. As a check that we read the curves correctly, the borrow rate they imply matches the interest Drift actually charged to within about $1\%$ on both markets, including through a change in the shape of the curve in mid 2025.

Measuring a supplier's share needs individual balances, and Drift records these only from 1 December 2025, as the balance an account holds after each deposit or withdrawal. Between two such records we grow the balance by the interest the market paid. This gives the balance of every account that deposited or withdrew during the four months to 1 April 2026, and dividing by the market's total deposits gives its share. Accounts with one owner count as one supplier, and so does a vault, since one manager decides its deposits. A \emph{lender} is a supplier that did not trade perpetual futures in the window. Table~\ref{tab:drift} gives the results.

\begin{table}[htbp]
\centering
\caption{Exposure on two Drift spot markets. Shares of hours, with every hour weighted equally. A lender is a supplier with no perpetual futures settlement in the window. Lender rows are lower bounds, and the rows for all suppliers are approximate.}
\label{tab:drift}
\small
\begin{tabular}{@{}lrr@{}}
\toprule
 & SOL & USDC \\
\midrule
\multicolumn{3}{@{}l}{\emph{11 November 2024 to 1 April 2026}} \\
Above the kink & $5.4\%$ & $36.6\%$ \\
$\alpha_{\mathrm{safe}}$ below $5\%$ & $0.3\%$ & $3.0\%$ \\
Median $\alpha_{\mathrm{safe}}$ above the kink & $9.9\%$ & $8.4\%$ \\
\midrule
\multicolumn{3}{@{}l}{\emph{1 December 2025 to 1 April 2026}} \\
Above the kink & $12.4\%$ & $26.0\%$ \\
A lender exposed, at least & $11.4\%$ & $7.9\%$ \\
Any supplier exposed, approximately & $11.4\%$ & $11.6\%$ \\
\bottomrule
\end{tabular}
\end{table}

\begin{figure}[htbp]
\centering
\includegraphics[width=\textwidth]{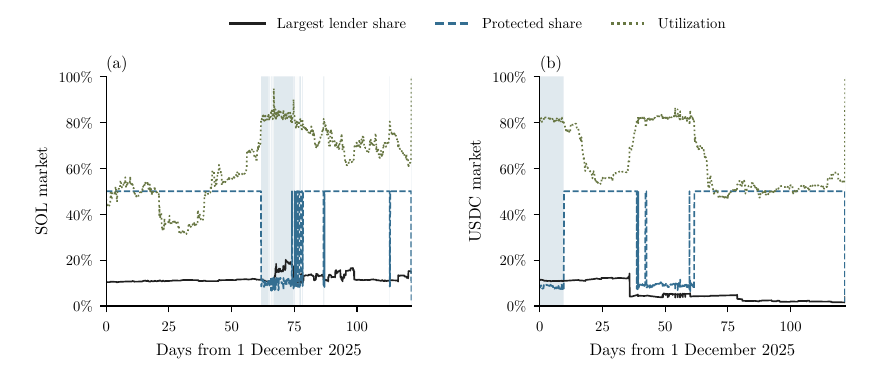}
\caption{Hourly utilization, protected share $\alpha_{\mathrm{safe}}$ and the largest lender's share of supply on Drift from 1 December 2025, for (a) SOL and (b) USDC. A lender is exposed wherever its share lies above the protected share, and those hours are shaded.}
\label{fig:drift}
\end{figure}

In at least one hour in nine on SOL and one in thirteen on USDC, a lender held more of the pool than the curve protected, so depositing more would have lowered that lender's income. On SOL one lender was exposed for $219$ hours and held up to $20\%$ of supply during them, and on USDC one was exposed for $230$ hours while holding up to $11.5\%$. Figure~\ref{fig:drift} shows the largest lender's share against the protected share. Drift's curve rarely protected less than $5\%$ of the pool and never less than $2\%$. The exposure came from lenders holding $10$ to $20\%$ of supply while the protected share above the kink sat near $8$ to $10\%$. Whom a curve protects matters more than how steep it is.

Three points govern how far these counts can be trusted. First, accounts that did not deposit or withdraw during the window leave no record, so the rebuilt suppliers cover $57\%$ of SOL deposits and $62\%$ of USDC deposits at the median hour. Adding the missing accounts back could only reveal more exposed suppliers, never remove one, and combining accounts by owner can only raise a share, so the counts understate exposure. Second, the rebuilt balances are accurate. For exposed lenders, the balance we carry forward matches the next recorded balance to within $1\%$ in $99.7\%$ of cases on SOL and in every case on USDC. Third, suppliers that also trade perpetual futures are harder to track, because trading gains and losses change their balances without a deposit record. Across all exposed suppliers, traders included, the match rate falls to $99.5\%$ on SOL and $95\%$ on USDC. The lender counts are therefore lower bounds, and the counts that include traders are approximate.

The counts say nothing about what these suppliers actually did. Depositing more would have lowered their income, but whether any of them held back for that reason is a separate question. Suppliers that also trade perpetual futures hold deposits partly as collateral and may not size them by interest, which is why the lender rows lead. The analysis code and a pseudonymized snapshot of the data, with accounts numbered instead of named, are in the repository.

\section{The Admissible Family}
\label{sec:family}

The condition $\alpha(1+\eta)\leq1$ describes a supplier rather than a curve. A protocol cannot control $\alpha$, but it can decide whose deposits to keep profitable. Write $\bar\alpha(U)$ for the largest supplier share the protocol intends to protect when the resulting utilization is $U$, and call this a \emph{protection profile}. Imposing the condition at $\alpha=\bar\alpha(U)$ gives $1+\eta\leq1/\bar\alpha$, a ceiling on elasticity that we call the \emph{elasticity envelope} $E$ induced by the profile. The ceiling also covers every smaller supplier, because $\alpha(1+\eta)$ grows with $\alpha$,
\begin{equation}
\boxed{
\eta(U)\leq E(U)
\quad\text{where}\quad
E(U)=\frac{1-\bar\alpha(U)}{\bar\alpha(U)}
\qquad\forall U
}
\label{eq:envelope}
\end{equation}
Protecting more of the pool forces a lower envelope: a target of $\bar\alpha=\tfrac12$ allows nothing beyond unit elasticity, $\bar\alpha=5\%$ allows $19$, and as $\bar\alpha\to0$ the envelope grows without bound and the constraint disappears.

During a deposit the supplier's share only grows. Under a constant $\bar\alpha$ it is therefore enough to check the final share. A profile that varies with $U$ must be checked along the path.

Since elasticity is a logarithmic derivative (Equation~\ref{eq:elasticity}), writing the constraint as
\begin{equation}
\frac{d\ln R_b}{d\ln U}\leq E(U)
\end{equation}
and integrating it bounds the curve itself. We say that $R_b$ \emph{honors} the profile $\bar\alpha$ at $U$ when $\eta(U)\leq E(U)$. Equivalently, the curve's own safe share $\alpha_{\mathrm{safe}}(U)=1/(1+\eta(U))$ is at least $\bar\alpha(U)$. So $\alpha_{\mathrm{safe}}$ is what a given curve protects, and $\bar\alpha$ is what the protocol sets out to protect.

\begin{theorem}[The admissible family]
\label{thm:family}
Let $0<U_a<U_b<1$, let $R_b$ be positive and locally absolutely continuous, and let $E(U)/U$ be integrable on the interval. Then $R_b$ honors the profile $\bar\alpha$ almost everywhere on $[U_a,U_b]$ if and only if
\begin{equation}
\boxed{
R_b(U)\leq R_b(U_a)\exp\left(\int_{U_a}^{U}\frac{E(v)}{v}\,dv\right)
}
\label{eq:family}
\end{equation}
holds for every $U\in[U_a,U_b]$ and on every subinterval. Equality throughout characterizes the \emph{boundary curve} of the profile, the solution of $UR_b'(U)=E(U)R_b(U)$ anchored at $(U_a,R_b(U_a))$. Endpoint statements follow by limits where those limits exist.
\end{theorem}
\begin{proof}
If $\eta\leq E$, then $R_b'/R_b\leq E/U$. The left side is the derivative of $\ln R_b$, so integrating from $U_a$ to $U$ and exponentiating gives the bound. Conversely, apply the bound to a short interval $[U,U+h]$, take logarithms and divide by $h$. The left side becomes the average slope of $\ln R_b$ over the interval, and the right side the average of $E(v)/v$. As $h\to0$ they tend to $R_b'(U)/R_b(U)$ and $E(U)/U$, which gives $\eta(U)\leq E(U)$. These limits exist at every point outside a set of zero length, which is why the result holds almost everywhere. For the piecewise smooth curves used here they exist inside every segment.
\end{proof}

The next sections develop a constant share target and protection for every fresh depositor, and Appendix~\ref{app:examples} illustrates their boundary curves.

\section{Fixed Supplier Share Protection}
\label{sec:fixed}

The simplest policy is to protect the same maximum supplier share at every utilization. The profile is then a constant $\bar\alpha$, so the envelope in Equation~\ref{eq:envelope} is a constant too,
\begin{equation}
\boxed{
K=\frac{1-\bar\alpha}{\bar\alpha}
}
\end{equation}
and the design rule is $\eta(U)\leq K$ everywhere.

With $E\equiv K$ in Theorem~\ref{thm:family}, the boundary anchored at the kink $(O,R_o)$ is
\begin{equation}
\boxed{
R_K(U)=R_o\left(\frac{U}{O}\right)^K
}
\end{equation}
This power law is the equality case of Equation~\ref{eq:family}, so its elasticity is identically $K$.

\subsection{The Endpoint Impossibility Bound}
\label{sec:endpoint}

A rate curve must still reach $R_{\max}$ at full utilization, and fixed endpoints therefore limit how small the worst case elasticity can be. Integrating Equation~\ref{eq:elasticity} from $O$ to $1$ shows that any positive differentiable curve connecting $(O,R_o)$ to $(1,R_{\max})$ satisfies
\begin{equation}
\ln\frac{R_{\max}}{R_o}
=\int_O^1 \eta(U)\,d\ln U
\label{eq:endpoint}
\end{equation}
so if $\eta\leq K$ everywhere then $\ln(R_{\max}/R_o)\leq K\ln(1/O)$.

\begin{theorem}[Endpoint impossibility bound]
\label{thm:kstar}
Any positive, absolutely continuous borrow rate curve with endpoints $(O,R_o)$ and $(1,R_{\max})$ has essential supremum of elasticity at least
\begin{equation}
\boxed{
K^*=\frac{\ln(R_{\max}/R_o)}{\ln(1/O)}
}
\end{equation}
so an elasticity ceiling $K<K^*$ is infeasible whatever the shape of the curve.
\end{theorem}

The bound is tight, since the power law with $K=K^*$ reaches $R_{\max}$ exactly at $U=1$. Appendix~\ref{sec:smooth} shows that with a continuous derivative no curve attains $K^*$.

Since protected share is $1/(1+K)$, the endpoints also cap the guarantee a protocol can offer without moving them,
\begin{equation}
\boxed{
\bar\alpha_{\max}=\frac{1}{1+K^*}
}
\end{equation}
Optimal utilization, the rate at optimal utilization, the maximum rate, and the supplier protection target therefore cannot be chosen independently of one another. For the illustrative inputs $O=80\%$, $R_o=5\%$ and $R_{\max}=100\%$ this gives $K^*\approx13.43$ and $\bar\alpha_{\max}\approx6.93\%$, so no curve through those endpoints guarantees protection to a supplier holding more than about $7\%$ of pool supply.

\section{The Fresh Depositor Guarantee}
\label{sec:fresh}

A different policy is to protect anyone arriving with no existing position, whatever the state of the market when they arrive. Suppose such a supplier deposits $x$ into a pool with initial supply $S_0$ and fixed borrowings $B$, and write
\begin{equation}
u_0=\frac{B}{S_0}
\qquad\text{and}\qquad
u=\frac{B}{S_0+x}
\end{equation}
for utilization before and after. Their final pool share depends only on those two utilizations,
\begin{equation}
\alpha=\frac{x}{S_0+x}=1-\frac{u}{u_0}
\end{equation}
and substituting into $\alpha(1+\eta)\leq1$ gives the exact condition along a deposit path beginning at $u_0$,
\begin{equation}
\boxed{(u_0-u)R_b'(u)\leq R_b(u)}
\end{equation}

For a nondecreasing rate the left side grows with $u_0$, so the most demanding starting point is $u_0=1$, where $\alpha=1-u$. Protecting every fresh depositor from every feasible start is therefore just another protection profile, and everything in Section~\ref{sec:family} applies unchanged.

\begin{theorem}[Global fresh depositor condition]
\label{thm:fresh}
Protecting a fresh depositor entering from any initial utilization is the protection profile
\begin{equation}
\boxed{\bar\alpha(U)=1-U}
\end{equation}
whose envelope is
\begin{equation}
\boxed{E(U)=\frac{U}{1-U}}
\end{equation}
so the design rule is $\eta(u)\leq u/(1-u)$, equivalently $(1-u)R_b'(u)\leq R_b(u)$, and the boundary curve is
\begin{equation}
\boxed{
R_b(U)=R_b(U_a)\,\frac{1-U_a}{1-U}
}
\end{equation}
\end{theorem}
\begin{proof}
Substituting $\bar\alpha=1-U$ into Equation~\ref{eq:envelope} gives $E=U/(1-U)$, and $\eta\leq E$ rearranges to $(1-u)R_b'\leq R_b$. Since $E(v)/v=1/(1-v)$, the integral in Equation~\ref{eq:family} evaluates to $\ln\big((1-U_a)/(1-U)\big)$, which gives the stated boundary.
\end{proof}

\begin{figure}[htbp]
\centering
\includegraphics[width=\textwidth]{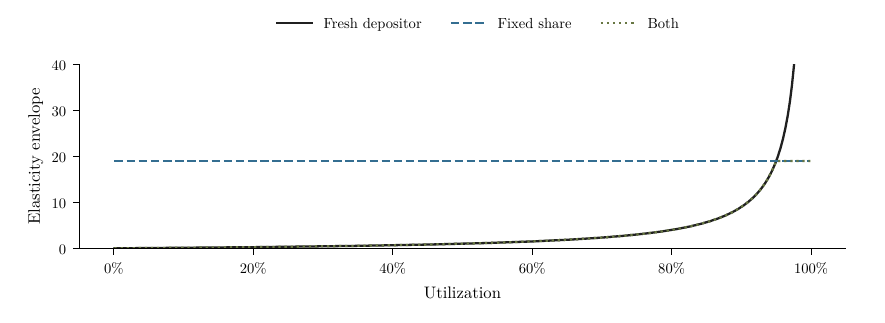}
\caption{Elasticity envelopes of the fresh depositor guarantee and a $5\%$ flat target. Imposing both guarantees takes the smaller envelope.}
\label{fig:envelope}
\end{figure}

The profile protects half the pool at $50\%$ utilization, a tenth of it at $90\%$, and only a hundredth at $99\%$. Promising little near full utilization is what lets the rule tolerate a steep emergency branch, and it is also the rule's main limitation, since it promises nothing to the larger suppliers a pool short of liquidity most needs. Figure~\ref{fig:envelope} compares its envelope with that of a $5\%$ flat target. The two cross at $U=K/(1+K)$, which is $95\%$ for $K=19$. Below that point the fresh depositor rule is the stricter one and above it the flat target is, so neither guarantee implies the other.

One further consequence separates this profile from a flat one. A fresh depositor guarantee cannot be offered by a curve through the origin. Applying Theorem~\ref{thm:fresh} on $[U_a,U]$ and letting $U_a\to0$ gives $R_b(U)\leq R_b(0)/(1-U)$, that is
\begin{equation}
R_b(0)\geq R_b(U)\,(1-U)
\qquad\forall U<1
\end{equation}
so the base rate is bounded below by the whole curve and in particular is strictly positive. The standard below kink ramp fails for exactly this reason: being proportional to $U$ it has $\eta\equiv1$, while the envelope $U/(1-U)$ is below $1$ for every $U<\tfrac12$. The example curve in Appendix~\ref{app:examples}, Table~\ref{tab:cavey-example}, starts at a positive rate.

\subsection{Piecewise Linear Form}

Onchain curves are piecewise linear, and the guarantee takes an exact form on them. Take a nondecreasing segment from $(u_L,R_L)$ to $(u_H,R_H)$ with slope
\begin{equation}
m=\frac{R_H-R_L}{u_H-u_L}
\end{equation}
so that $R_b(u)=R_L+m(u-u_L)$ inside it. The condition $\eta\leq u/(1-u)$ is equivalent to $m(1-u)\leq R_b(u)$, and since $R_b(u)/(1-u)$ increases along a nondecreasing segment, the binding point is the left endpoint. The segment is admissible exactly when $0\leq m\leq R_L/(1-u_L)$, and substituting the slope turns this into a rule on adjacent kinks alone,
\begin{equation}
\boxed{
\frac{R_H}{R_L}
\leq
1+\frac{u_H-u_L}{1-u_L}
}
\label{eq:adjacent-rate-rule}
\end{equation}
If every segment of a continuous piecewise linear curve satisfies Equation~\ref{eq:adjacent-rate-rule}, income stays nondecreasing even for a deposit that crosses several segments, and no condition linking one segment to the next is required. A flat clamp after the final kink preserves the condition as well.

Appendix~\ref{app:examples} gives an admissible curve and follows a fresh deposit across its kinks.

\section{A Minimal Change Construction}
\label{sec:minimal}

Return to a flat target and suppose a protocol already runs a single kink model
\begin{equation}
R_{\mathrm{old}}(U)=
\begin{cases}
R_o\dfrac{U}{O}, & U\leq O\\[5pt]
R_o+m(U-O), & U>O
\end{cases}
\qquad
m=\frac{R_{\max}-R_o}{1-O}
\end{equation}
Assume $K\geq1$ and the upper branch has negative intercept, $R_o-mO<0$, as in the illustrative curve. Below the optimum $\eta=1$ and nothing needs to change. Above it $\eta_{\mathrm{old}}(U)=Um/(R_o+m(U-O))$ is largest at the kink. For $K\geq K^*$, let $U_c$ be the first utilization above $O$ at which the power law boundary catches the original branch,
\begin{equation}
R_o\left(\frac{U_c}{O}\right)^K
=
R_o+m(U_c-O)
\end{equation}
For a feasible $K$ the natural construction follows the boundary only over that interval,
\begin{equation}
\boxed{
R_{\mathrm{new}}(U)=
\begin{cases}
R_o\dfrac{U}{O}, & U\leq O\\[6pt]
R_o\left(\dfrac{U}{O}\right)^K, & O<U<U_c\\[7pt]
R_o+m(U-O), & U_c\leq U\leq1
\end{cases}
}
\end{equation}
preserving the curve below the kink and after the rejoin. The replacement can extend into a region where the original elasticity is already below $K$, because the power law branch runs below the original and must still catch up with it. If the original curve is admissible everywhere, no replacement interval is needed. In particular, once $K$ exceeds the largest original elasticity the construction returns the original curve, so it is better read as a constraint on an existing interest rate model than as a replacement for it.

\begin{figure}[htbp]
\centering
\includegraphics[width=\textwidth]{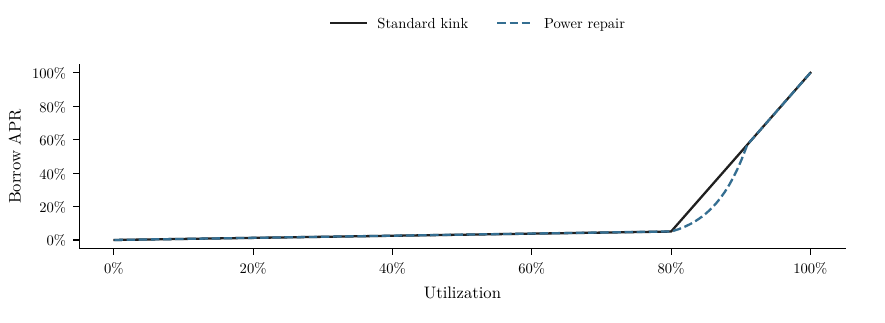}
\caption{Borrow APR before and after the power law repair with $K=19$. The repaired branch rejoins at approximately $90.92\%$ utilization.}
\label{fig:borrow-curves}
\end{figure}

Figure~\ref{fig:borrow-curves} compares the borrow rates for $O=80\%$, $R_o=5\%$, $R_{\max}=100\%$ and $K=19$. The branch changes the path between the fixed endpoints while keeping both the kink rate and the maximum rate.

\begin{figure}[htbp]
\centering
\includegraphics[width=\textwidth]{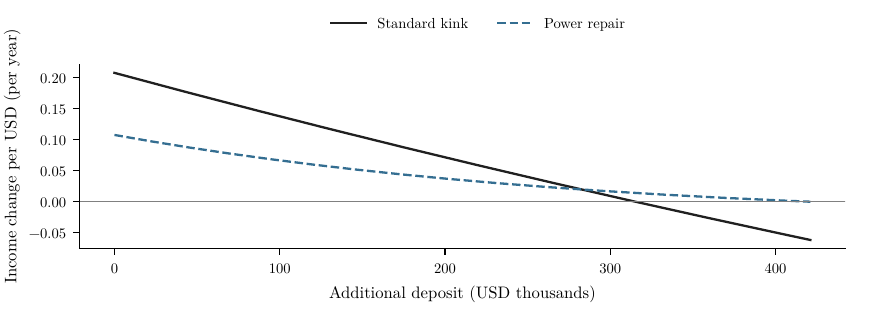}
\caption{Marginal income of a supplier that starts with \$100{,}000 in a \$10 million pool with \$8.5 million of fixed borrowing and deposits until it owns $5\%$. Under the original curve it turns negative, while under the power repair it stays at or above zero.}
\label{fig:partial-supplier-income}
\end{figure}

Over the replaced interval this curve charges borrowers less and pays suppliers less, but a deposit can no longer lower the income of a supplier within the protected share. A supplier that starts with \$100{,}000 in a \$10 million pool at $85\%$ utilization and deposits until it owns $5\%$ eventually loses income under the original curve, while under the repair its income never falls, as Figure~\ref{fig:partial-supplier-income} shows. Preserving existing income levels requires a different pricing choice, considered next.

\section{Repairs and Price Levels}
\label{sec:multi}

Consider a USDT market, with parameters drawn from a deployed configuration, having a $0.5\%$ minimum borrow APR, an $80\%$ main kink at $5.25\%$ APR, and a maximum borrow APR of $90\%$. Further kinks lie at $85\%$, $90\%$, $95\%$, $99\%$ and $99.5\%$ utilization, with rates specified in Appendix~\ref{app:multi}. The example lets us distinguish the protection a curve provides from the price levels chosen to provide it.

\begin{figure}[htbp]
\centering
\includegraphics[width=\textwidth]{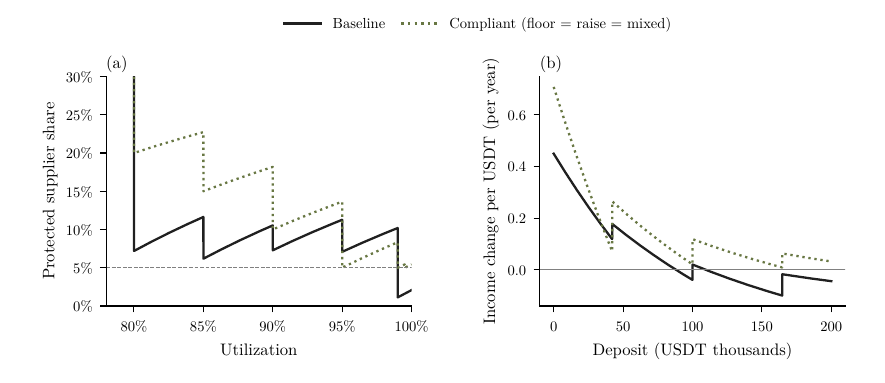}
\caption{Both guarantees on the fixed kinks. (a) Actual protected share, with the $5\%$ target marked. (b) Marginal whole position income for a fresh depositor entering a pool with $1$ million USDT supply and $990{,}000$ USDT of fixed borrowing. The floor, raise only and mixed curves coincide here. Their marginal income stays nonnegative across the crossed kinks, but the baseline's does not.}
\label{fig:multi-marginal}
\end{figure}

To protect both fresh depositors and existing suppliers up to a fixed share, impose the stronger promise at each utilization:
\begin{equation}
\bar\alpha_{\mathrm{both}}(U)=\max\{\bar\alpha,1-U\}
\qquad
E_{\mathrm{both}}(U)=\min\left\{K,\frac{U}{1-U}\right\}
\end{equation}
A power law region alone does not establish both guarantees, because every retained segment must also respect the tighter envelope. Figure~\ref{fig:multi-marginal} shows the protected share and the marginal income of a fresh depositor under both guarantees.

\begin{figure}[htbp]
\centering
\includegraphics[width=\textwidth]{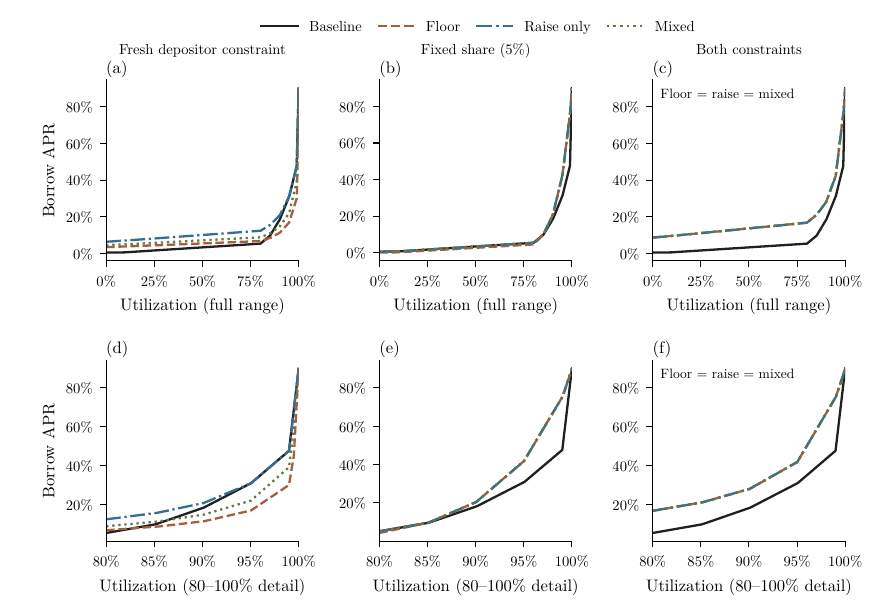}
\caption{Baseline, compliant floor, raise only repair and mixed repair, at fixed kinks and $90\%$ top APR. Columns impose the fresh depositor guarantee, the $5\%$ fixed share guarantee, and both. Panels (a--c) show $0$--$100\%$ utilization, and panels (d--f) show the same curves enlarged over $80$--$100\%$, with no change in parameters or scenario. Under both constraints the floor is already above the baseline, so the three repairs coincide.}
\label{fig:repair-family}
\end{figure}

With kinks and maximum APR fixed, the segment constraints place a floor under preceding rates. Each guarantee caps the ratio between adjacent kink rates, $r_{i+1}\leq c_i r_i$. For the fresh depositor guarantee $c_i$ is the right side of Equation~\ref{eq:adjacent-rate-rule}, and Appendix~\ref{app:multi} gives it for each guarantee. The floor is obtained by propagating the final rate backwards as $\ell_i=\ell_{i+1}/c_i$. Moving every rate to the floor can lower rates that were already acceptable. A repair that preserves them instead sets
\begin{equation}
r_n^+=r_n^0
\qquad
r_i^+=\max\left\{r_i^0,\frac{r_{i+1}^+}{c_i}\right\}
\label{eq:raise-only}
\end{equation}
where $r_i^0$ denotes the baseline. This is the smallest admissible repair that never lowers a rate, so it preserves supplier income under the same principal and utilization history at a cost to borrowers. Figure~\ref{fig:repair-family} compares these choices with a mixed repair that permits increases and decreases while minimizing weighted changes to kink rates. The recursions, objective and feasibility conditions are derived in Appendix~\ref{app:multi}.

On the original kinks, both guarantees with $5\%$ fixed share protection and a $90\%$ maximum require a base APR of at least $8.68\%$, and Appendix~\ref{app:multi} shows how far curved regions between the kinks lower this bound. A power law branch that only lowers rates after the $99\%$ kink reaches just $57.65\%$ at full utilization, so keeping the $90\%$ maximum requires raising rates earlier, as Appendix~\ref{sec:analytic} shows.

\subsection{Income Over a Holding Period}
\label{sec:earnings}

Whether a depositor earns more over a holding period depends on the rates they encounter and the capital they hold at each point, rather than on the sign of the deposit derivative alone. Levels, unlike signs, depend on the reserve factor $f$ that Section~\ref{sec:model} set aside. For prescribed principal $b(t)$ and utilization $U(t)$, accumulated interest is
\begin{equation}
A(H)=\int_0^H(1-f)b(t)U(t)R_b(U(t))\,dt
\end{equation}
where time is measured in years. Comparing two curves on identical histories gives
\begin{equation}
\Delta A(H)=\int_0^H(1-f)b(t)U(t)
\left[R_1(U(t))-R_0(U(t))\right]dt
\end{equation}
A raise only repair therefore cannot reduce accumulated supplier interest under the same history, and a lower only repair cannot increase it. A mixed change depends on where the depositor spends time and how much capital is present there.

As in Section~\ref{sec:model}, borrowings do not respond to the new rates. The comparisons replay the same utilization path under every curve, and predicting the response would need a demand model such as those of~\cite{baude-optimal,agilerate}. The examples use a $5\%$ reserve factor, pay interest out rather than compounding it, and omit other fees and token rounding.

\begin{figure}[htbp]
\centering
\includegraphics[width=\textwidth]{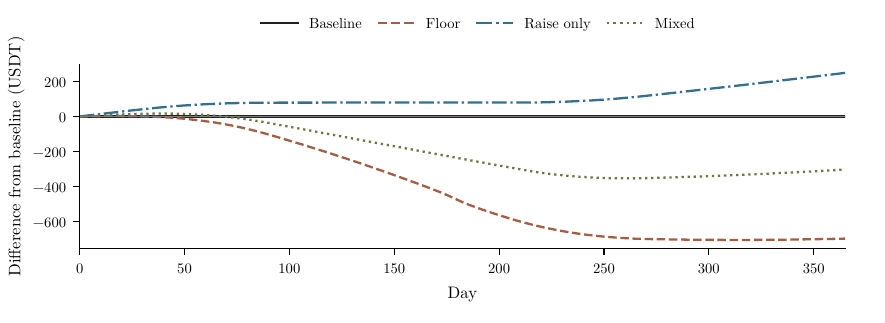}
\caption{Accumulated interest relative to the baseline under the fresh depositor guarantee alone, on the single deposit history. Principal is excluded.}
\label{fig:fresh-lifecycle}
\end{figure}

First take a $10{,}000$ USDT depositor in a pool with $1$ million USDT supply. They hold for one year while utilization moves linearly through $80\%$, $95\%$, $99.5\%$, $85\%$ and $80\%$ on days $0$, $90$, $180$, $270$ and $365$. Under the fresh depositor guarantee alone, the floor pays about $1{,}287$ USDT, the raise only repair $2{,}231$, and the mixed repair $1{,}680$, against $1{,}983$ under the baseline. Under both guarantees the three repairs coincide and pay about $3{,}165$ USDT. Figure~\ref{fig:fresh-lifecycle} shows how the fresh depositor repairs separate over the year. All three repairs obey the same guarantee and keep the same maximum. Their different earnings come from the choice of rate levels.

Higher supplier earnings on an unchanged history are funded by borrower payments. Since this depositor owns $1\%$ of the pool, each additional unit of aggregate borrower interest contributes $0.0095$ units to the depositor after the reserve factor. The guarantee governs how income responds to a deposit. The price level is a separate choice, and it decides who pays.

Appendix~\ref{app:histories} considers repeated deposits, stronger protection targets and different utilization exposures.

\section{Checking the Constraint Onchain}
\label{sec:onchain}

Onchain implementations represent rate curves with linear segments rather than evaluating fractional powers or exponentials, and the flat share constraint can be checked directly in that form. For a segment from $(u_i,r_i)$ to $(u_{i+1},r_{i+1})$ with slope $m_i$,
\begin{equation}
R_b(U)=r_i+m_i(U-u_i)
\qquad
\eta(U)=\frac{Um_i}{R_b(U)}
\end{equation}
and elasticity is monotone across a positive linear segment, so checking both endpoints bounds it over the whole segment. Writing $\Delta r=r_{i+1}-r_i$ and $\Delta U=u_{i+1}-u_i$, the condition at each endpoint $(U,r)$ of the segment can be put in a form with no division at all,
\begin{equation}
\boxed{
U\Delta r\leq Kr\Delta U
}
\end{equation}
which suits fixed point integer arithmetic and imposes the elasticity requirement on an otherwise standard piecewise linear interest rate model. It belongs in parameter validation rather than in the rate computation, so the hot path is untouched.

The check applies to the curve actually implemented. As Appendix~\ref{app:implementation} shows, replacing an analytic boundary with straight chords can break its ceiling, and integer evaluation needs headroom for rounding.

\section{Conclusion}

The condition $\alpha(1+\eta)\leq1$ decides whether a deposit can lower a supplier's own income. When it fails, the supplier gains by withdrawing, at the moment the pool needs supply, and on two Drift markets real lenders were in that position in roughly one hour in ten. A protocol turns the condition into a rule on curves by choosing whom to protect. Protecting a fixed share gives the power law boundary and the endpoint bound $K^*$, and protecting every fresh depositor gives the hyperbolic boundary and the adjacent rate rule of Equation~\ref{eq:adjacent-rate-rule}. Neither guarantee contains the other. The guarantee does not set the price level, so a protocol adopting it should state its fixed endpoints, allowed kink changes and protection target together.

Several questions remain open: how borrowers respond to repricing, the welfare effects of the different repairs, monotonicity at the transaction level after token rounding, smooth curve families beyond the one built in Appendix~\ref{sec:smooth}, and whether exposed suppliers on live markets act on their exposure.

\clearpage
\appendix
\setcounter{figure}{0}
\renewcommand{\thefigure}{A\arabic{figure}}
\renewcommand{\theHfigure}{appendix.\arabic{figure}}
\section{Boundary Curves and Deposit Examples}
\label{app:examples}

\begin{figure}[htbp]
\centering
\includegraphics[width=\textwidth]{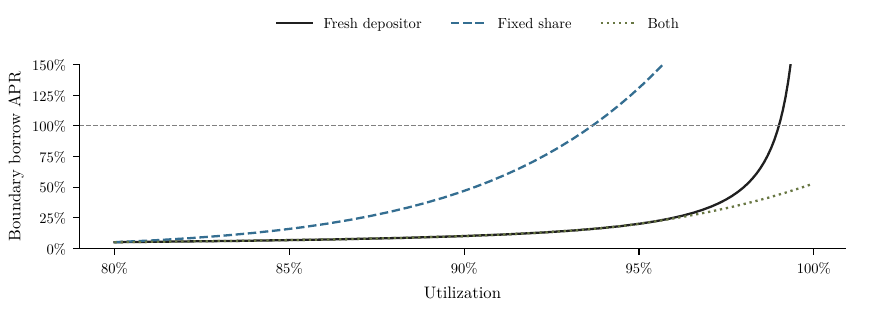}
\caption{Boundary curves anchored at $(80\%,5\%)$, with a $100\%$ rate reference. The combined boundary uses the smaller envelope and cannot reach that reference.}
\label{fig:boundary-curves}
\end{figure}

Figure~\ref{fig:boundary-curves} draws the boundary curves of both guarantees and their combination.

\subsection{A Fresh Depositor Crossing Several Kinks}

\begin{figure}[htbp]
\centering
\includegraphics[width=\textwidth]{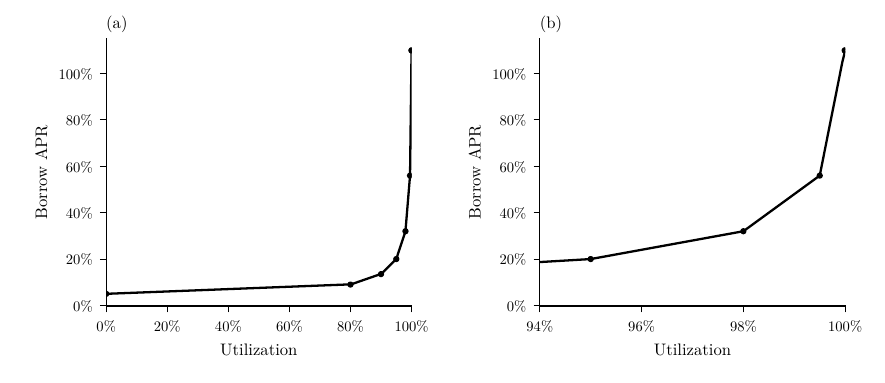}
\caption{A piecewise linear curve through points at $0\%$, $80\%$, $90\%$, $95\%$, $98\%$, $99.5\%$ and $99.99\%$ utilization and borrow APRs of $5\%$, $9\%$, $13.5\%$, $20\%$, $32\%$, $56\%$ and $110\%$, followed by a flat extension to the endpoint at $100\%$ utilization (also marked). Every segment satisfies Equation~\ref{eq:adjacent-rate-rule}, and the nonzero rate at the origin is forced by the base rate bound.}
\label{fig:deposit-monotonic-curve}
\end{figure}

The following curve satisfies the adjacent rate rule from Section~\ref{sec:fresh} and retains an increasing emergency branch near full utilization. Table~\ref{tab:cavey-example} checks each segment, and Figure~\ref{fig:deposit-monotonic-curve} plots the curve.

\begin{table}[htbp]
\centering
\caption{A piecewise linear borrow rate curve satisfying deposit monotonicity.}
\label{tab:cavey-example}
\begin{tabular}{@{}rrrr@{}}
\toprule
Segment & Rate ratio & Maximum ratio & Result \\
\midrule
$0\%\!\to\!80\%$       & 1.800 & 1.800 & tight \\
$80\%\!\to\!90\%$     & 1.500 & 1.500 & tight \\
$90\%\!\to\!95\%$     & 1.481 & 1.500 & valid \\
$95\%\!\to\!98\%$     & 1.600 & 1.600 & tight \\
$98\%\!\to\!99.5\%$   & 1.750 & 1.750 & tight \\
$99.5\%\!\to\!99.99\%$ & 1.964 & 1.980 & valid \\
\bottomrule
\end{tabular}
\end{table}

\begin{figure}[htbp]
\centering
\includegraphics[width=\textwidth]{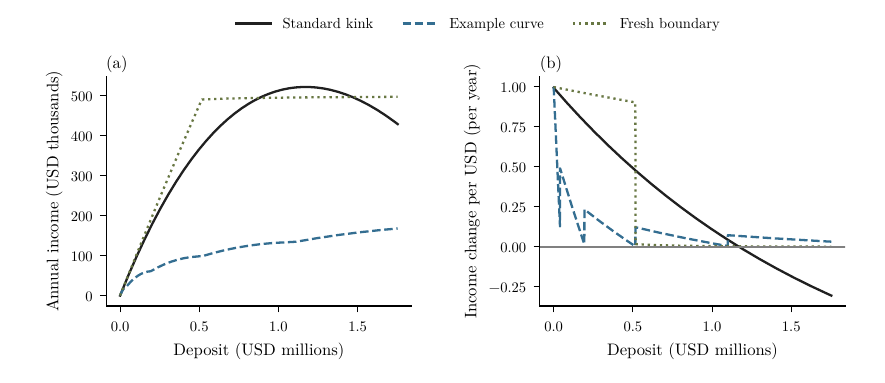}
\caption{(a) Annualized income and (b) marginal income along the fresh depositor path. The standard kink has zero base rate. The example curve of Table~\ref{tab:cavey-example} and the clamped fresh boundary have $5\%$ base rates. The example curve reaches $110\%$ maximum APR, while the other two reach $100\%$.}
\label{fig:fresh-income}
\end{figure}

Figure~\ref{fig:fresh-income} shows the guarantee along one deposit path. A fresh depositor arrives at a pool with \$10 million of supply and \$9.99 million of fixed borrowing, and deposits until utilization reaches $85\%$. The path crosses the kinks at $99.5\%$, $98\%$, $95\%$ and $90\%$. Under Table~\ref{tab:cavey-example}, income rises over the whole path. Under the standard kinked curve of Section~\ref{sec:failure}, income peaks and then falls, and its marginal dollar turns negative at a deposit of about \$1.17 million.

The curve in Table~\ref{tab:cavey-example} ends the path at about \$168{,}000 a year against \$428{,}000 for the kinked one, because it charges $20\%$ at $95\%$ utilization where the rule would allow $100\%$. The third curve takes equality in Theorem~\ref{thm:fresh} from a base rate $r_0=5\%$, giving $R_b(U)=\min\{r_0/(1-U),100\%\}$. It charges about $33\%$ at $85\%$ utilization against the standard curve's $29\%$, reaches its maximum at $95\%$, and ends the path at about \$497{,}000 a year.

Below the clamp the boundary's marginal income is
\begin{equation}
\frac{dI}{dx}=\frac{r_0u^2(1-u_0)}{u_0(1-u)^2}
\end{equation}
which is zero only for the most demanding start $u_0=1$ and slightly positive for the illustrated start $u_0=0.999$.

\section{Constructions for Several Kinks}
\label{app:multi}

For the market of Section~\ref{sec:multi}, the rate below the main kink is $\max\{0.5\%,5.25\%\,U/0.8\}$. Above it, straight segments connect the points in Table~\ref{tab:multi}. The parameters are drawn from a deployed configuration. They are not measurements of live market activity or a proposed parameter update. All fixed kink recursions below use the complete grid: $0\%$, the exact floor crossing $U=0.005(0.8)/0.0525\approx7.619\%$, and the seven utilizations in Table~\ref{tab:multi}.

\begin{table}[htbp]
\centering
\caption{The illustrative multiple kink rate curve. The floor meets the initial ramp at approximately $7.619\%$ utilization.}
\label{tab:multi}
\begin{tabular}{@{}rrrrrrrr@{}}
\toprule
Utilization & $80\%$ & $85\%$ & $90\%$ & $95\%$ & $99\%$ & $99.5\%$ & $100\%$ \\
\midrule
Borrow APR & $5.25\%$ & $9.4875\%$ & $17.9625\%$ & $30.675\%$ & $47.625\%$ & $68.8125\%$ & $90\%$ \\
\bottomrule
\end{tabular}
\end{table}

\subsection{Segment Caps}

Both guarantees together use the smaller of the two envelopes, as in Section~\ref{sec:multi}, so every segment must satisfy whichever envelope is tighter there. On a linear segment $[u_i,u_{i+1}]$, write $\Delta u_i=u_{i+1}-u_i$. The fresh depositor cap on $r_{i+1}/r_i$ is
\begin{equation}
c_i^{\mathrm{fresh}}=1+\frac{\Delta u_i}{1-u_i}
\end{equation}
Checking the fixed share condition at both endpoints gives
\begin{equation}
c_i^{\mathrm{fixed}}=\min\left\{
1+\frac{K\Delta u_i}{u_i},\
\frac{u_{i+1}}{u_{i+1}-K\Delta u_i}
\right\}
\end{equation}
where a term imposes no restriction if $u_i=0$ in the first denominator or the second denominator is nonpositive. Taking $c_i=\min\{c_i^{\mathrm{fresh}},c_i^{\mathrm{fixed}}\}$ imposes both guarantees, with the complete segment constraint $r_i\leq r_{i+1}\leq c_i r_i$.

\subsection{Rate Floors and Repair Choices}

If a market must still reach its existing maximum, limiting the increase on each segment places a lower bound on the rates leading up to it. For fixed kinks and final rate $r_n$, this bound follows by setting $\ell_n=r_n$ and working backwards:
\begin{equation}
\ell_i=\frac{\ell_{i+1}}{c_i}
\end{equation}
An unbounded cap gives a zero lower bound. Every feasible curve with these kinks and top rate must lie above the resulting floor, but moving all rates down to it can reduce supplier income even where the baseline was already compliant. A protocol that wants to preserve that income can instead require each repaired rate to remain at least as high as its baseline value.

With baseline rates $r_i^0$, the least such repair is the raise only recursion of Equation~\ref{eq:raise-only}. Because the baseline is nondecreasing, the backwards recursion satisfies both segment inequalities. Any admissible curve above the baseline must meet the same backwards bounds, making this the pointwise smallest repair that preserves existing rates. Supplier income cannot fall under an unchanged market history, although the higher rates increase borrower costs.

The opposite choice fixes the base and never raises a baseline rate:
\begin{equation}
r_0^-=r_0^0,\qquad
r_{i+1}^-=\min\{r_{i+1}^0,c_i r_i^-\}
\end{equation}
This is the greatest admissible curve below the baseline, but its top rate is not fixed. Keeping the original top while never raising a rate is infeasible whenever the baseline falls below the corresponding fixed top floor.

Allowing both increases and decreases gives a third pricing choice. The mixed repair in the figures minimizes $\sum_i w_i(r_i-r_i^0)^2$ over feasible kink rates with the top held fixed, using trapezoidal utilization weights $w_i$. This objective measures changes to the kink parameters. It does not optimize borrower or supplier welfare, nor is it the exact integral of squared curve error.

\begin{figure}[htbp]
\centering
\includegraphics[width=\textwidth]{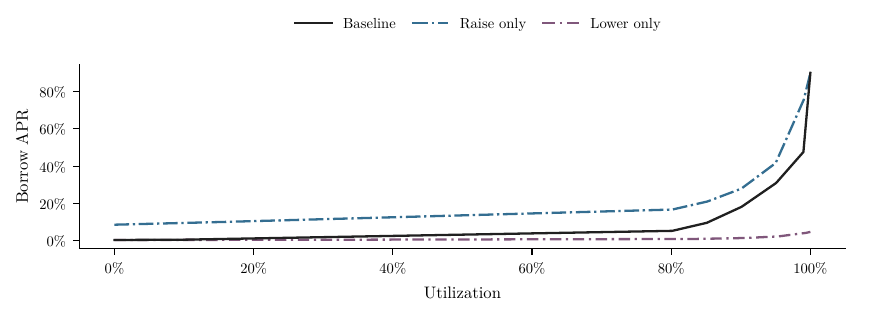}
\caption{Rates under both guarantees. Raise only keeps the $90\%$ top and coincides with the floor and mixed repair. Lower only never raises a rate, so it must lower the maximum.}
\label{fig:repair-direction}
\end{figure}

For the fresh depositor rule, the fixed top floor has base APR $3.47\%$, while preserving every baseline rate requires $6.39\%$. Under both guarantees the minimum base is $8.68\%$ on these kinks, and the floor, raise only and mixed repairs coincide. A repair that never raises the baseline instead reaches only $12.04\%$ maximum APR under the fresh rule and $4.82\%$ under both, rather than $90\%$. Figure~\ref{fig:repair-direction} compares the two directions under both guarantees.

\subsection{Power Law Regions on a Multiple Kink Curve}
\label{sec:analytic}

With several kinks, a rate lowering power branch may fail to catch the original curve before full utilization even when the main kink and maximum satisfy the global endpoint bound. Retaining low rates at intermediate kinks can leave too little utilization range for the final increase, so preserving the maximum may require raising rates before the offending segment.

For the constructions in this subsection, assume a continuous, nondecreasing piecewise linear baseline with strictly positive base rate, as in the illustrative curve. This ensures the running infimum minorant remains positive. For the fixed share envelope, a baseline with a positive slope first segment through the origin is also permitted when $K\geq1$. When $K<1$, its running infimum minorant is identically zero and is outside the positive rate class.

These replacement regions can be constructed by measuring the baseline relative to the fastest rate increase allowed by the guarantee. Let $F$ be a positive boundary function on $U>0$, with $F'/F=E(U)/U$, and write $q_0(U)=R_0(U)/F(U)$. Admissibility requires $R/F$ to be nonincreasing: once the rate has fallen behind its boundary, it cannot catch up by exceeding the envelope later. The largest admissible curve below the baseline and the smallest one above it are therefore
\begin{equation}
\boxed{
\begin{aligned}
R_-(U)&=F(U)\inf_{0<v\leq U}q_0(v)\\
R_+(U)&=F(U)\sup_{U\leq v\leq1}q_0(v)
\end{aligned}
}
\end{equation}
Values at zero are taken by continuity. The first construction preserves as much borrower pricing as possible without raising any rate. The second preserves supplier pricing by raising only what the constraint requires.

\begin{proof}
The envelope condition is equivalent to $R/F$ being nonincreasing. Any admissible minorant therefore satisfies $R(U)/F(U)\leq q_0(v)$ for every $v\leq U$, giving the first bound. The running infimum is itself nonincreasing and meets that bound. The majorant follows by the same argument with future points and a running supremum. On a piecewise linear nondecreasing baseline, the resulting curve follows either the baseline or a constant multiple of $F$. Both are nondecreasing, and the construction is continuous at contacts. Thus the bounds also belong to the rate class assumed here.
\end{proof}

\begin{figure}[htbp]
\centering
\includegraphics[width=\textwidth]{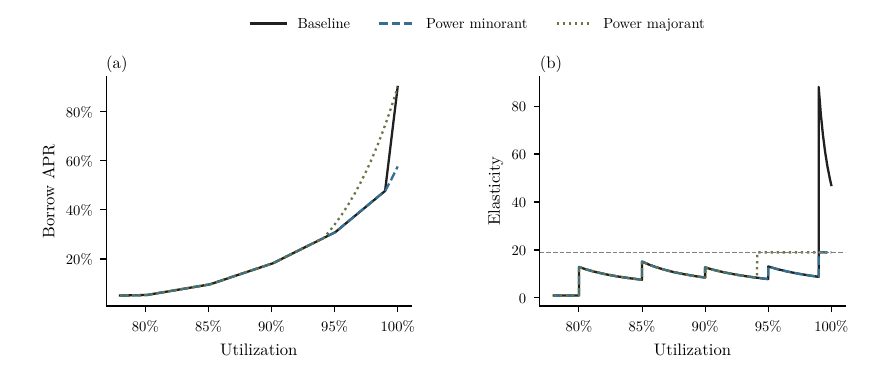}
\caption{Analytic fixed share repairs with $K=19$: (a) borrow APR and (b) elasticity. The power minorant never raises rates but misses the original maximum. The power majorant preserves the maximum and raises the rates needed to reach it. The horizontal line marks the elasticity ceiling.}
\label{fig:analytic-power}
\end{figure}

For fixed share protection, $F(U)=U^K$, so the replacement intervals are power laws. On the illustrative multiple kink curve with $K=19$, the minorant first departs at $99\%$ utilization and reaches only $57.65\%$ APR at full utilization. It cannot retain the $90\%$ maximum. The majorant retains that maximum by raising rates earlier, while still preserving the original base and $80\%$ kink rate. Figure~\ref{fig:analytic-power} shows both curves and their elasticity. The formulas include interior extrema of $q_0$ on each original segment, since checking the values at the kinks alone is not sufficient for the running supremum.

\begin{figure}[htbp]
\centering
\includegraphics[width=\textwidth]{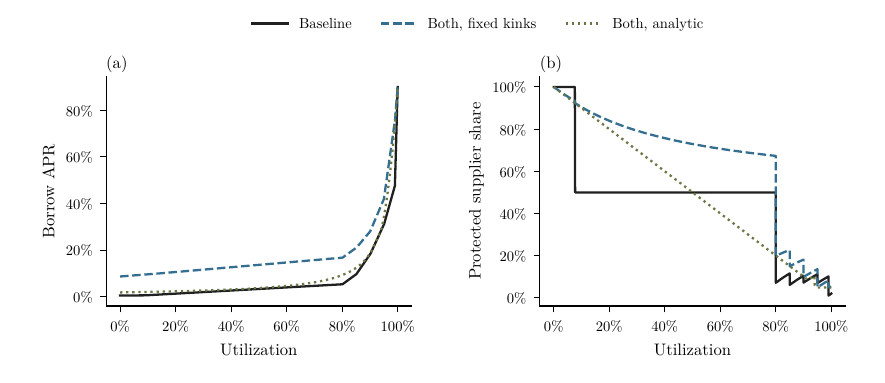}
\caption{Both guarantees, with $5\%$ fixed share protection. The fixed kink and analytic curves are the least repairs nowhere below the same baseline in their respective classes. (a) Borrow APR. (b) Actual protected share. The analytic curve can place new contacts and use curved regions, so it is not restricted to the original kink grid.}
\label{fig:analytic-combined}
\end{figure}

To impose both guarantees, let $t=K/(K+1)$ and normalize the combined boundary at the origin:
\begin{equation}
F(U)=
\begin{cases}
(1-U)^{-1},&0\leq U\leq t\\[3pt]
(K+1)(U/t)^K,&t<U\leq1
\end{cases}
\end{equation}
The same construction then uses hyperbolic regions below $t$ and power law regions above it, retaining the baseline wherever contact permits. This is distinct from the fixed kink linear implementation. The combined analytic majorant has base APR approximately $1.85\%$ and keeps the $90\%$ maximum, compared with $8.68\%$ for the fixed kink repair. Figure~\ref{fig:analytic-combined} compares the two.

\subsection{The Cost of Keeping the Kinks}

The combined envelope has a finite budget even though the fresh depositor envelope alone does not. With no restriction on kink placement, its fixed top base bound is
\begin{equation}
\boxed{
R_b(0)\geq\frac{R_{\max}}{F(1)}
=\frac{R_{\max}t^K}{K+1}
}
\end{equation}
For $K=19$ and $R_{\max}=90\%$, this is $1.70\%$. It is attained by a scaled combined boundary. That curve need not lie above the baseline. Preserving the baseline raises the bound to the $1.85\%$ found above.

\begin{figure}[htbp]
\centering
\includegraphics[width=\textwidth]{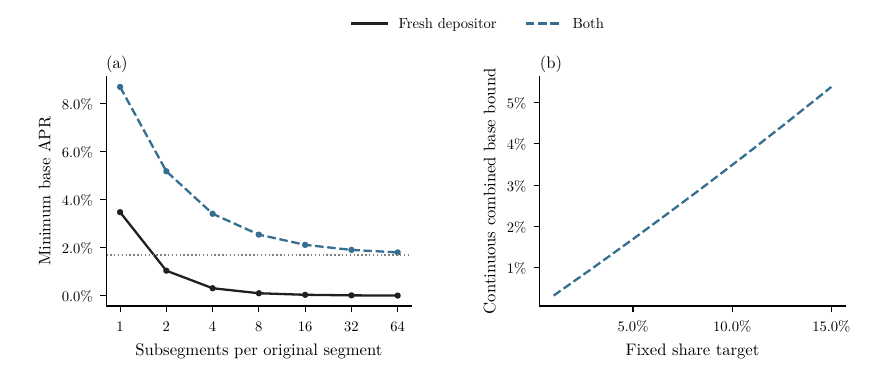}
\caption{(a) Minimum base APR at a fixed $90\%$ maximum as every original interval is subdivided equally. The dotted reference is the continuous combined bound, not a floor for the fresh only rule. (b) The continuous combined base bound as the fixed share target varies. These floors do not additionally require staying above the baseline.}
\label{fig:knot-sensitivity}
\end{figure}

Figure~\ref{fig:knot-sensitivity} subdivides each original segment and recalculates the linear floor. At $64$ subsegments per original segment, its combined base falls from $8.68\%$ to $1.80\%$, approaching the continuous bound. Under the fresh rule alone it falls towards zero as more of the rate increase can be concentrated near full utilization. A finite set of kinks and the continuous admissible family are different design spaces, and their bounds should not be interchanged.

\section{Additional Interest Scenarios}
\label{app:histories}

The histories below extend Section~\ref{sec:earnings}. They use the same constant $5\%$ reserve factor, pay interest out rather than compounding it, and prescribe identical principal and utilization paths for each rate curve.

\begin{figure}[htbp]
\centering
\includegraphics[width=\textwidth]{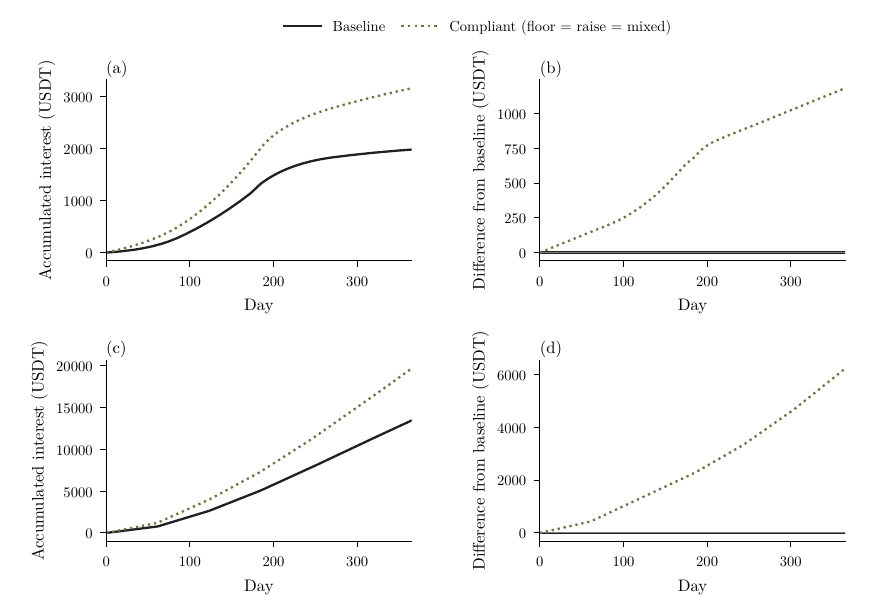}
\caption{Both guarantees on fixed kinks. Top row: the single deposit history. Bottom row: repeated additions into fixed borrowing. Left: accumulated interest. Right: difference from baseline. All curves use identical principal and utilization histories, a $5\%$ reserve factor and no compounding. Exit at the end returns principal separately.}
\label{fig:both-lifecycle}
\end{figure}

Figure~\ref{fig:both-lifecycle} repeats the single deposit history of Section~\ref{sec:earnings} under both guarantees and adds a second history. It begins with a $10{,}000$ USDT depositor in a pool holding $1$ million USDT against $990{,}000$ borrowed. The depositor adds $20{,}000$ on each of days $60$, $120$, $180$, $240$ and $300$, while borrowing remains fixed. Annual interest is about $19{,}667$ USDT under the combined repair and $13{,}453$ under the baseline. Ownership eventually exceeds $5\%$, so the local guarantee still has to be read from the protection profile rather than inferred from the label of the curve.

\begin{figure}[htbp]
\centering
\includegraphics[width=\textwidth]{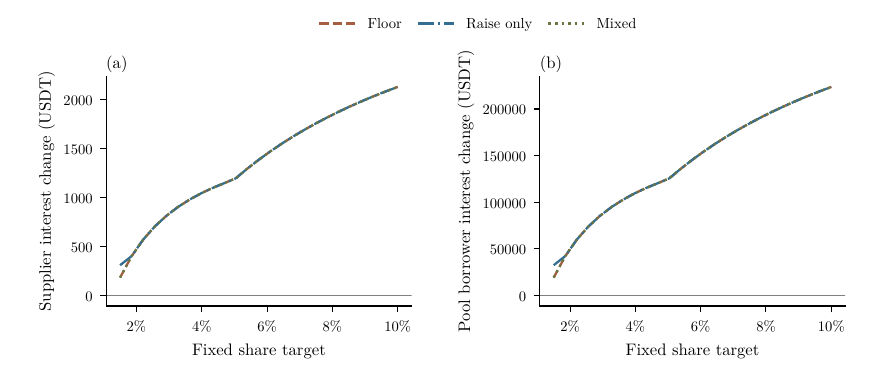}
\caption{Both guarantees as the fixed share target varies. (a) Annual supplier interest change and (b) aggregate borrower interest change, on the single deposit history. The repairs coincide over much of this range. With fixed $1\%$ ownership and a $5\%$ reserve factor, panel (a) is $0.0095$ times panel (b).}
\label{fig:protection-cost}
\end{figure}

Figure~\ref{fig:protection-cost} follows the transfer from borrowers to suppliers as the fixed share target varies, keeping the fresh guarantee, the original kinks and the maximum.

\begin{figure}[htbp]
\centering
\includegraphics[width=\textwidth]{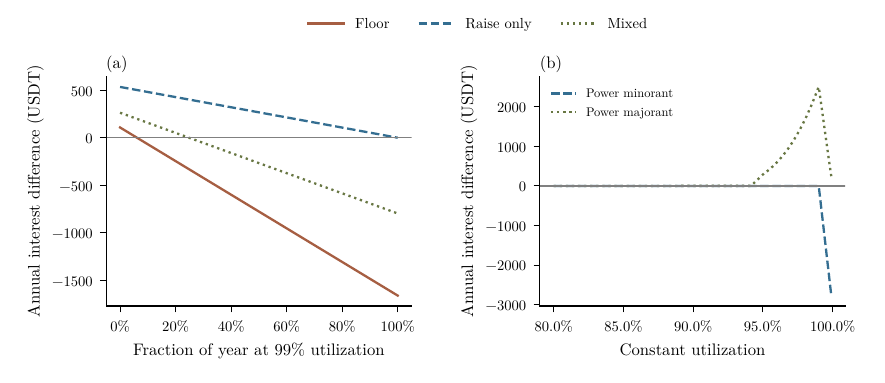}
\caption{Sensitivity to utilization exposure. (a) Fresh only fixed kink repairs: a year is split between $80\%$ and $99\%$ utilization, with the high utilization fraction on the horizontal axis. The top legend applies to this panel. (b) Fixed share analytic power repairs: annual interest differences at a constant utilization, with a separate legend. Both panels hold principal at $10{,}000$ USDT and use a $5\%$ reserve factor.}
\label{fig:exposure}
\end{figure}

To test dependence on utilization exposure, Figure~\ref{fig:exposure} allocates a year between $80\%$ and $99\%$ utilization for the same $10{,}000$ USDT principal. The fresh only floor can gain at low utilization and lose at high utilization, whereas the raise only repair cannot lose under any such mix. The analytic power minorant and majorant have the corresponding sign restrictions.

\section{Discretization and Numerical Checks}
\label{app:implementation}

\begin{figure}[htbp]
\centering
\includegraphics[width=\textwidth]{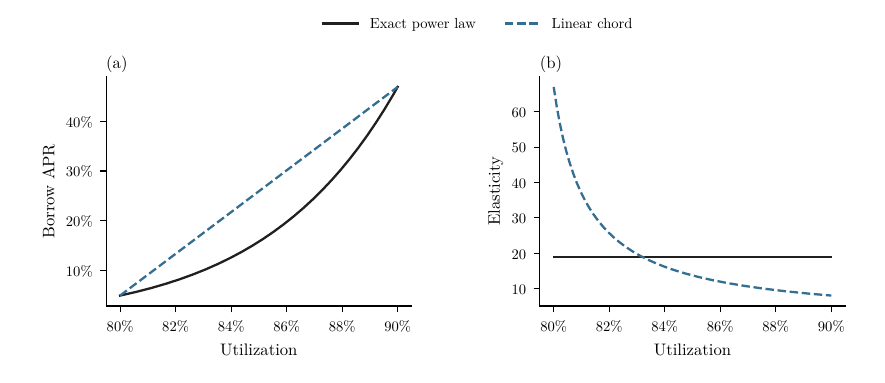}
\caption{(a) A constant elasticity power law and its chord between $80\%$ and $90\%$ utilization. (b) Their elasticities. Matching endpoint rates does not preserve the ceiling between the endpoints.}
\label{fig:discretization}
\end{figure}

One caveat applies to any discretization of a smooth boundary. Sampling the constant elasticity boundary at two points replaces it by a chord whose slope is the average of the slopes it spans, so the chord is steeper than the boundary at the segment's left endpoint. That endpoint is where elasticity binds, since the chord of a convex curve has negative intercept and $\eta$ therefore decreases across the segment, and the sampled curve is more elastic there than the boundary it was drawn from. Sampling $R_o(U/O)^{19}$ at $80\%$ and $90\%$, for instance, gives a segment whose elasticity at $80\%$ is about $67$. Figure~\ref{fig:discretization} shows this chord. A ceiling imposed on a sampled curve has to be chosen with considerable headroom, or the kinks placed close enough together that the chords track the boundary.

Integer evaluation adds small steps where no derivative exists. In the multiple kink example, the integer formula at all $1{,}000{,}001$ utilization ticks stays within $10^{-6}$ APR of the rational curve, but that does not prove monotonicity after token and interest rounding, so an implementation needs rounding headroom and finite increment checks at its own precision.

\section{Smooth Constructions}
\label{sec:smooth}

A protocol may also want the advertised protected share to vary continuously with utilization. This requires a continuous elasticity profile and imposes an additional cost beyond the deposit monotonicity constraint.

\begin{figure}[htbp]
\centering
\includegraphics[width=\textwidth]{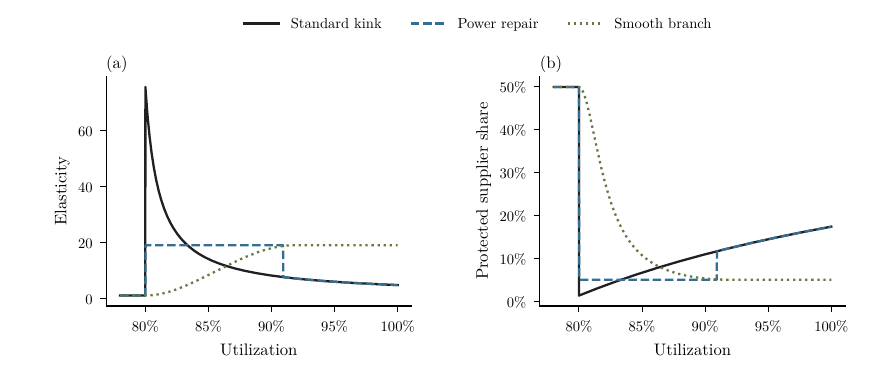}
\caption{(a) Elasticity and (b) protected share for the standard kink, the minimal power repair and a smooth branch. The repaired curves satisfy $K=19$, but the standard kink does not. The power repair jumps in protected share, while the smooth branch changes continuously.}
\label{fig:protected-share}
\end{figure}

The minimal change construction is continuous in the rate but not in its derivative: below the kink the slope is $R_o/O$ while the power branch leaves at $KR_o/O$, so unless $K=1$ the two disagree, and the same mismatch occurs in reverse at $U_c$. This affects the guarantee a protocol can advertise, because the protected share $\alpha_{\mathrm{safe}}=1/(1+\eta)$ is computed from elasticity and inherits every discontinuity of it. Under the minimal change curve $\alpha_{\mathrm{safe}}$ falls from $\tfrac12$ to $1/(1+K)$ at a single point, so a supplier holding $30\%$ of the pool is protected at $U=O-\varepsilon$ and unprotected at $U=O+\varepsilon$, and the guarantee the protocol advertises jumps while the state variable it depends on moves continuously. Figure~\ref{fig:protected-share} shows the jump next to the smooth branch built below.

The integral of Theorem~\ref{thm:family} also builds curves instead of only bounding them. If $\eta$ is continuous on $[U_a,U_b]$ then
\begin{equation}
R_b(U)=R_a\exp\left(\int_{U_a}^{U}\frac{\eta(v)}{v}\,dv\right)
\end{equation}
is positive, continuously differentiable, and has elasticity exactly $\eta$. Conversely every positive $C^1$ curve through $(U_a,R_a)$ is of this form with $\eta$ its own elasticity, and a piecewise continuous $\eta$ gives the same correspondence away from its jumps. Prescribing elasticity is therefore equivalent to prescribing the rate, and the construction below works in that direction, because it makes the protected share continuous.

Removing the jump means asking elasticity itself to vary continuously. Choose a continuous profile $\eta$ on $[O,1]$ that starts at $\eta(O)=\eta_0$ and stays below a ceiling $K$, and recover the rate from it by the integral above with $U_a=O$ and $R_a=R_o$. Then $R_b'=\eta R_b/U$ is continuous, the branch is $C^1$, and matching $\eta_0$ at the kink makes the assembled curve $C^1$ across the join. Appendix~\ref{app:smoothstep} carries out the construction explicitly for one such family.

The ceiling cannot be taken all the way down to the endpoint bound, however.

\begin{theorem}[$C^1$ makes $K^*$ unattainable]
\label{thm:strict}
Let $R_b\in C^1([O,1])$ be positive with $R_b(O)=R_o$ and $R_b(1)=R_{\max}$, and let its elasticity satisfy $\eta(O)=\eta_0<K^*$. Then
\begin{equation}
\boxed{\sup_{U\in[O,1]}\eta(U)>K^*}
\end{equation}
so a ceiling $K=K^*$ is infeasible under $C^1$, while every $K>K^*$ remains feasible.
\end{theorem}
\begin{proof}
In $x=\ln(U/O)$ with $L=\ln(1/O)$, Equation~\ref{eq:endpoint} reads $\int_0^L\eta\,dx=K^*L$, so $K^*$ is the average elasticity the curve must achieve. Continuity with $\eta(0)=\eta_0<K^*$ holds $\eta$ below $K^*$ over an interval of positive length, and an average of $K^*$ then forces $\eta>K^*$ somewhere later. Feasibility for $K>K^*$ follows from the construction in Appendix~\ref{app:smoothstep}.
\end{proof}

Theorem~\ref{thm:kstar} makes $K^*$ the minimum worst case elasticity and Theorem~\ref{thm:strict} makes it a minimum that is approached without being reached, since the curve attaining it is the pure power law, which is exactly the one that gives up $C^1$ at the kink.

The power repair is the greatest admissible curve below the single kink baseline and is $C^0$, with a protected share that drops at the kink. A continuous elasticity profile gives $C^1$ and a protected share that slides, but only by spending elasticity headroom above $K^*$: for the illustrative inputs, a ceiling of $K=19$ buys a transition spread from $80\%$ to $91.9\%$ utilization. Both leave $O$, $R_o$ and $R_{\max}$ where they were.

\subsection{An Explicit \texorpdfstring{$C^1$}{C1} Branch}
\label{app:smoothstep}

For a concrete construction, we use the following family to compute the headroom required by smoothing.

Work in $x=\ln(U/O)\in[0,L]$ with $L=\ln(1/O)$ and $A=\ln(R_{\max}/R_o)=K^*L$, and let $s(t)=3t^2-2t^3$ be the standard smoothstep, which has $s(0)=0$, $s(1)=1$, $s'(0)=s'(1)=0$ and $\int_0^1 s=\tfrac12$. For $0<w\leq L$, prescribe elasticity as a ramp that finishes within the interval, followed by a plateau,
\begin{equation}
\boxed{
\eta(x)=
\begin{cases}
\eta_0+(\eta_{\max}-\eta_0)\,s(x/w), & 0\leq x\leq w\\[3pt]
\eta_{\max}, & w<x\leq L
\end{cases}
}
\end{equation}
and recover $R_b$ from it by the integral construction. Taking $\eta_0=1$ matches the ramp below the kink, so the curve is $C^1$ there, and the vanishing endpoint slopes of $s$ make $\eta$ itself $C^1$ and hence $R_b\in C^2$.

The plateau is not a free parameter, since the branch must still arrive at $R_{\max}$.

\begin{proposition}[The cost of smoothing]
\label{prop:width}
The endpoint condition $\int_0^L\eta\,dx=A$ determines the plateau uniquely,
\begin{equation}
\boxed{
\eta_{\max}(w)=\frac{A-\eta_0w/2}{L-w/2}
\qquad 0\leq w\leq L
}
\end{equation}
with $\eta_{\max}(0)=K^*$ and, whenever $\eta_0<K^*$,
\begin{equation}
\frac{d\eta_{\max}}{dw}=\frac{L(K^*-\eta_0)}{2\,(L-w/2)^2}>0
\end{equation}
so $\eta_{\max}(w)>K^*$ for every $w>0$ in this domain. The value at $w=0$ denotes the nonsmooth limit. For a width in $[0,L]$, the branch respects a ceiling $K>\eta_0$ if and only if
\begin{equation}
\boxed{
w\leq\frac{2L\,(K-K^*)}{K-\eta_0}
}
\end{equation}
\end{proposition}
\begin{proof}
Integrating the profile, $\int_0^w\eta\,dx=\eta_0w+(\eta_{\max}-\eta_0)w\int_0^1s=w(\eta_0+\eta_{\max})/2$, so
\begin{equation}
\int_0^L\eta\,dx=\frac{w(\eta_0+\eta_{\max})}{2}+\eta_{\max}(L-w)=\eta_{\max}\left(L-\frac{w}{2}\right)+\frac{\eta_0w}{2}
\end{equation}
Setting this equal to $A$ and solving gives $\eta_{\max}$, and differentiating gives the stated derivative, whose sign is that of $K^*-\eta_0$ because $A=K^*L$. Imposing $\eta_{\max}(w)\leq K$ and rearranging yields the width bound.
\end{proof}

Proposition~\ref{prop:width} is the quantitative form of Theorem~\ref{thm:strict}: the plateau meets $K^*$ only in the degenerate case $w=0$, which is the power law and is not $C^1$. The rate is available in closed form as well, with $\int_0^t s=t^3-t^4/2$,
\begin{equation}
R_b(U)=R_o\exp H(x)
\qquad
H(x)=
\begin{cases}
\eta_0x+(\eta_{\max}-\eta_0)\,w\left[t^3-\tfrac{t^4}{2}\right]_{t=x/w}, & x\leq w\\[5pt]
\dfrac{w(\eta_0+\eta_{\max})}{2}+\eta_{\max}(x-w), & x>w
\end{cases}
\end{equation}
so evaluating it requires no quadrature.

\begin{figure}[htbp]
\centering
\includegraphics[width=\textwidth]{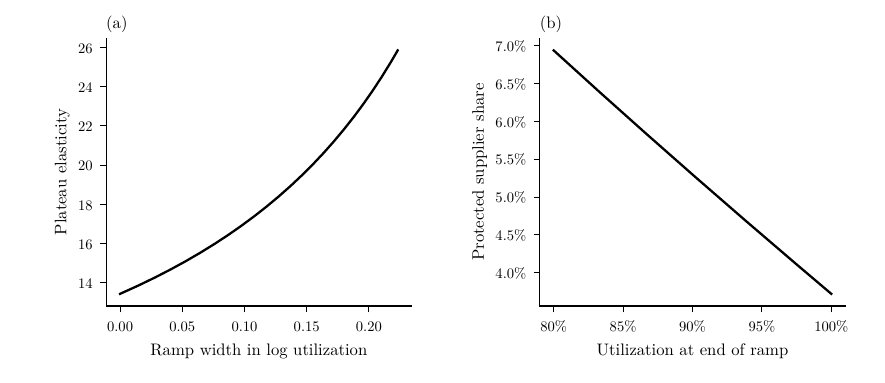}
\caption{(a) Plateau elasticity against ramp width, with $0\leq w\leq L$. (b) The corresponding protected share against the utilization where the ramp ends. The value $K^*$ is reached only in the nonsmooth limit $w=0$.}
\label{fig:smoothness-cost}
\end{figure}

Table~\ref{tab:smoothing} records the trade for the illustrative inputs, where $L=0.2231$, $A=2.9957$ and $K^*=13.43$. A wider ramp buys a gentler transition and costs protected share, and at $w=L$ the ramp fills the whole branch, the plateau is reached only at full utilization, and $\eta_{\max}=2K^*-\eta_0$, which equals $2K^*-1$ under the standing choice $\eta_0=1$. For a $5\%$ target, meaning $K=19$, the widest admissible ramp has $w=0.1382$ and carries elasticity from $1$ to $19$ between $80\%$ and $91.9\%$ utilization while still reaching $R_{\max}$ exactly at $U=1$. Figure~\ref{fig:smoothness-cost} plots the same trade over all widths.

\begin{table}[htbp]
\centering
\caption{Smoothing width against plateau elasticity, for $O=80\%$, $R_o=5\%$, $R_{\max}=100\%$ and $\eta_0=1$.}
\label{tab:smoothing}
\begin{tabular}{@{}rrrr@{}}
\toprule
Width $w$ & Ramp ends at & $\eta_{\max}$ & Protected share \\
\midrule
0.0000 & $80.00\%$ & 13.43 & $6.93\%$ \\
0.0500 & $84.10\%$ & 14.99 & $6.25\%$ \\
0.1000 & $88.41\%$ & 17.01 & $5.55\%$ \\
0.1382 & $91.86\%$ & 19.00 & $5.00\%$ \\
0.2231 & $100.00\%$ & 25.85 & $3.72\%$ \\
\bottomrule
\end{tabular}
\end{table}

\end{document}